\documentclass[12pt, a4paper]{article}
\usepackage[sc]{mathpazo}
\usepackage[left= 2.2cm, right = 2.2cm, top = 2.65cm, bottom = 2.65cm]{geometry}
\usepackage{graphicx, color, nicefrac, ragged2e}
\usepackage{amsmath, amssymb, bm}
\usepackage{relsize}
\usepackage{lmodern}
\usepackage{xcolor}

\newcommand{\comm}[1]{}
\usepackage{makecell}
\usepackage[flushleft]{threeparttable}
\usepackage{url}
\usepackage[numbers, sort&compress, comma]{natbib}
\definecolor{cobalt}{rgb}{0.0, 0.28, 0.67}
\definecolor{coolblack}{rgb}{0.0, 0.18, 0.39}
\definecolor{cordovan}{rgb}{0.54, 0.25, 0.27}
\definecolor{blue(pigment)}{rgb}{0.2, 0.2, 0.6}
\usepackage[%
colorlinks,bookmarksopen,bookmarksnumbered,citecolor=coolblack,
urlcolor=cordovan, linkcolor=cobalt]{hyperref}
\usepackage{subcaption}

\usepackage{algpseudocode,algorithm,algorithmicx}

\algrenewcommand\algorithmicrequire{\textbf{Precondition:}}
\algrenewcommand\algorithmicensure{\textbf{Postcondition:}}

\usepackage[toc,page]{appendix}
\usepackage{authblk}

\title{Piecewise survival models: a change-point analysis on herpes zoster associated pain data revisited and extended}
\author[1]{Dimitra Eleftheriou}
\author[2]{Dimitris Karlis}

\affil[1]{School of Mathematics and Statistics, University of Glasgow}
\affil[2]{Department of Statistics, Athens Univeristy of Economics and Business}

\providecommand{\keywords}[1]{\textbf{\textit{Keywords: }} #1}

\date{}

\begin{document}

\maketitle

\rule[0.35ex]{\linewidth}{1pt}

\begin{abstract}
For many diseases it is reasonable to assume that the hazard rate is not constant across time, but also that it changes in different time intervals. To capture this, we  work here with a piecewise survival model. One of the major problems in such piecewise models is to determine the time points of change of the hazard rate. From the practical point of view this can provide very important information as it may reflect changes in the progress of a disease. We present piecewise Weibull regression models with covariates. The time points where change occurs are assumed unknown and need to be estimated.  The  equality of hazard rates across the distinct phases is also examined to verify the exact number of phases. An example based on herpes zoster data has been used to demonstrate the usefulness of the developed methodology.
\\\\
\keywords{Piecewise regression modelling, change-point detection, simulation, simulated annealing, survival analysis, model selection, herpes zoster.}

\end{abstract}

\rule[0.35ex]{\linewidth}{1pt}

\section{Introduction}

Herpes Zoster is a viral disease characterized by a painful skin and vesicular rash. The occurrence of herpes zoster follows from recrudescence of latent varicella-zoster virus (VZV), which is reactivated and multiplied in the dorsal root of cranial nerve ganglia. After primary infection with varicella (i.e. chickenpox), the virus persists dormant in the central nervous system (CNS) of the infected patient causing damage to it. This neurological disease is a cause of considerable morbidity, especially in both elderly and immunosuppressed individuals \cite{dworkin2001epidemiology,wood2002understanding}. The
zoster-associated pain (ZAP) is the most problematic symptom of the disease as it can be highly debilitating, and can also limit the patient's productivity \cite{huff1988therapy}.

With regard to the classification of ZAP in 1994 by \citet{dworkin1994proposed},  a division of pain into three phases, measured in days, has been proposed as: \textit{acute herpetic neuralgia} [0,30), \textit{subacute herpetic neuralgia} [30,120), and \textit{postherpetic neuralgia (PHN} [120,$\infty$). In addition, the
same classification is verified in other research studies associated with herpes zoster pain data in order to model pain resolution \cite{arani2001phase,desmond2002clinical}. Prompt antiviral regiments in patients who suffer from VZV,  such as
Acyclovir (ACV) and Valaciclovir (VACV), can accelerate rash healing, decrease viral shedding and pain duration, and prevent or reduce the incidence of postherpetic neuralgia \cite{huff1988therapy,gnann2002clinical}. 
%The first effective
%orally administered drug was acyclovir, which, according to
%previous meta-analysis studies based on placebo-controlled trials,
%reduced significantly the duration of ZAP as compared to placebo
%drug \cite{wood1996oral,jackson1997effect,wood2002understanding}.
%Nevertheless, acyclovir did not appear to be equally efficient on
%the severity of PHN. On the other hand, Valaciclovir, an oral
%prodrug of acyclovir, proved to be equivalent and maybe more
%effective in both healing acute lesions and limiting postherpetic
%neuralgia \cite{beutner1995valaciclovir}. Hence, VACV is preferred
%to ACV because of its higher levels of antiviral activity.

In accordance to the aforementioned,  there are various and
slightly arbitrary definitions for these transition points for
which, in most of the existing relevant literature, there is an
assumption based on the clinical evolution of the disease rather
than data supported evidence \cite{beutner1995valaciclovir, wood1996oral, watson2010postherpetic}. It is therefore of considerable
interest to have a well-defined description and a valid separation
for the distinct phases of zoster pain.

Consequently, it is very important in medical research to consider a model that evaluates the effect of antiviral therapies on each phase, while controlling for the effect of the other phases as well as other covariates of interest.
%Common analytic methods,
%such as a Kaplan-Meier survival function or a Cox's model, have
%been used to assess the pain. However, such approaches do not
%adequately allow for phase comparison.
The arising challenge is to detect the transition times where the disease passes to the next stage and characterize ZAP, and also to answer the clinical questions; ``When acute becomes PHN pain?" and ``What are the
treatment effects for a specific phase?". This analysis of time-to-event data assumes different hazard rate over distinguished time intervals giving rise to piecewise survival models. From a practical point of view, such a change-point detection, where the curvature of the survival functions change, can provide significant information as it may reflect changes in the progress of the disease of interest.

To this direction, piecewise exponential models (PEM) which are able to model the hazard rate of ZAP throughout its distinct time periods have been proposed \cite{arani2001phase}. We proceed by considering the less developed piecewise Weibull models (PWM), as they provide  a richer family having as special case the constant per period hazard rate, but we also examine the turning points of the disease based on such models. Such points are considered as unknowns that have to be estimated. We obtain change-points estimates optimizing the likelihood function of the piecewise survival models. We also provide methodology to evaluate an antiviral therapy and figure out how the treatment effects can be compared for a specific phase of ZAP. However, the fit of piecewise models is not straightforward, especially if the points of change of the hazard rate are not known. In such cases standard approaches may fail, since for example some discontinuities may arise and hence, derivative based maximization may fail. We avoid this obstacle working with the Simulated Annealing (SA) optimization method, which does not use derivatives. The structure of PEM relies on the  approach of \citet{arani2001phase}, which examined the existence of three phases.

We propose an improved approach in the change-point analysis of herpes zoster by using Weibull models. We aim at contributing towards two directions. Firstly, by reexamining the herpes zoster data using a piecewise Weibull model to show that the model improves with respect to the piecewise exponential used so far,  providing data driven insights about the pain phases. Secondly, we estimate the change-points together with the parameters of the model via a Simulated Annealing method to allow for full inference.

The following sections are structured as follows. In Section 2 the statistical methods are described. The proposed piecewise regression models are extensively delineated for multiple and unknown change-points. In Section 3 we apply this methodology on Herpes Zoster associated pain data for both cases of known and unknown transition points. The available data set has already been analysed by \citet{beutner1995valaciclovir} as well as in a relevant analysis of ZAP conducted by \citet{desmond2002clinical}. Section 4 summarizes the main statistical and clinical findings of the paper. 

\section{Methodology}

\subsection{The Piecewise Weibull Model}

Among the class of parametric statistical models, the Weibull model is one of the most widely used. This model has been proved proper enough to describe a wide range of time-to-event data \cite{lawless2011statistical}. The hazard function for Weibull distribution is given by
\begin{equation}
h(t;\kappa, \lambda)=  \frac{\kappa}{\rho} \, \left( \frac{t}{\rho} \right) ^{\kappa-1} = \lambda \, \kappa \: t^{\kappa-1},
\end{equation}
\\*
where  $\kappa>0$ is the \textit{shape parameter}, $\rho>0$ is the \textit{scale parameter} of the distribution, and $\lambda= \rho^{-\kappa}$ is a reparameterization to simplify the notation.
Note that the hazard rate is constant over time if $\kappa=1$, and the Weibull model reduces to an exponential model. If $\kappa>1$, then the hazard increases as time increases, and decreases over time otherwise. The addition of this shape parameter offers to the Weibull model great flexibility. The mean, median and variance of a Weibull random variable $T$  can be expressed as $E(T)=\rho \: \Gamma(1+\frac{1}{\kappa})$,  $M(T)=\rho \:(\log 2)^{\frac{1}{\kappa}}$, $V(T)=\rho^2[\Gamma(1+\frac{2}{\rho})-(\Gamma(1+\frac{1}{\kappa}))^2]$, respectively. Furthermore, the \textit{survival} and \textit{probability
densities} are respectively  defined as: $S(t;\kappa, \lambda) =
\exp[-\lambda t^\kappa]$, $f(t;\kappa,\lambda)= - \, {S(t)}' =
\lambda \: \kappa \: t^{\kappa-1} \exp[-\lambda t^\kappa], \:\: t
\geqslant 0$.

The piecewise Weibull model (PWM) is a way to combine interpretability and more flexibility by lightening the strict assumption of the piecewise constant hazard model. In PWM case, the basic idea is the partition of the time scale into \textit{J} intervals. In the bibliography, there is a variety of forms for such a piecewise Weibull baseline hazard rate model.  According to \citet{casellas2007bayesian}, the first and simplest form of a piecewise Weibull baseline hazard rate model
%is given by $ h_0(t; \kappa, \lambda, \tau_k, \tau_{k+1})= \kappa \: \lambda \: t^{\kappa-1} %\exp{(\epsilon_{\kappa})}$, $\tau_{k} < t \leqslant \tau_{k+1} $, $k=1,...,J-1$ and %$\epsilon_{\kappa}$ denotes the vector of the regression parameters. With respect to this %model, 
each phase shares the same \textit{shape} and \textit{scale parameters}. However, we give an extended approach that allows variety among Weibull distribution's parameters. Consequently, the piecewise Weibull baseline hazard function is given by 
\begin{equation}
 h_0(t; \bm{\kappa},\bm{\lambda},\bm{\tau})= \begin{cases}
  \kappa_1 \: \lambda_1 \: t^{\kappa_1-1} & \text{if \: $0\leq t<\tau_1$} \\
   \kappa_2 \: \lambda_2 \: t^{\kappa_2-1} & \text{if \: $\tau_1 \leq t<\tau_2$} \\
  \,\,\, \vdots                                     \\
   \kappa_J \: \lambda_J \: t^{\kappa_J-1} & \text{if \: $\tau_{J-1}\leq t$ \,,}
  \end{cases} \\
  \label{h0pwm}
\end{equation}
\\*
where $\bm{\lambda ^{\mathsmaller T}} = (\lambda_1,\lambda_2,...,\lambda_J) = (\rho_1^{-\kappa_1}, \rho_2^{-\kappa_2},... \, ,\rho_J^{-\kappa_J})$. The number of change-points is $K=J-1$, and $\bm{\tau}=(\tau_1,\tau_2,...,\tau_{J-1})$ represents the change-point vector, while $\kappa_j$ and $\rho_j$ are the positive \textit{shape} and \textit{scale parameters} of \textit{j}-th interval, respectively. The corresponding survival is  
\begin{equation}
S(t; \bm{\kappa},\bm{\lambda},\bm{\tau})= \begin{cases}
 \exp[ - \lambda_1 \, t^{\kappa_1}], & \text{if \: $0\leq t<\tau_1$} \\
 \exp[ - \lambda_1 \, \tau_1^{\kappa_1} - \lambda_2 \, (t^{\kappa_2}-\tau_1^{\kappa_2})], & \text{if \: $\tau_1 \leq t<\tau_2$} \\
  \,\,\, \vdots                                     \\
  \exp[ - \lambda_1 \tau_1^{\kappa_1} - \lambda_2(\tau_2^{\kappa_2}-\tau_1^{\kappa_2}) - ... - \lambda_J(t^{\kappa_J}-\tau_{J-1}^{\kappa_J})], & \text{if \: $\tau_{J-1}\leq t$ \,.}
  \end{cases} \\
\end{equation}

Other formulations of piecewise Weibull models are analysed in \cite{demiris2015survival,li2014likelihood}, which are known as poly-Weibull models and the baseline hazard function arises as the sum of the \textit{J} independent Weibull-type hazards.

The heterogeneity of the population is stated by the vector of $p$
covariates  $\bm{x_{i}}$ for the \textit{i}-th individual. %Incorporating covariates implies a proportional hazards model, which splits time and covariates into two components. 
In general, the form of piecewise Weibull instantaneous hazard of $i$-th individual, whose the observed survival time belongs to the $j$-th stage of ZAP is defined as
%\begin{equation}
%\begin{split}
%h_{ij}(t; \bm{x_i},\bm{\theta})=  h_0(t; \bm{\kappa},\bm{\lambda},\bm{\tau}) \, \exp(\bm{\beta^{\mathsmaller T}_j} \, \bm{x_i}) , \: \: \: \tau_j\leq t<\tau_{j+1}
%\end{split}
%\end{equation}
\begin{equation}
h_{ij}(t; \bm{x_i},\bm{\theta})=\kappa_j \, \lambda_j \, t^{\kappa_j-1} =  \kappa_j \, \exp(\bm{\beta^{\mathsmaller T}_{j}} \bm{x_i}) \, t^{\kappa_j-1} , \:\:\:\: \tau_j\leq t<\tau_{j+1} \,\, ,
\end{equation}
where $\bm{\theta}=(\bm{\kappa},\bm{\tau},\bm{\beta})$ includes the unknown parameters of interest, which are involved into the model. These are the \textit{shape} parameters $\bm{\kappa}$, the regression coefficients $\bm{\beta}$ which constitute the \textit{scale parameters} and the change-points $\bm{\tau}$, respectively. The log-likelihood function for PWM in any stage is given by
\begin{equation}
\ell(\bm{\theta}) = \log \, L(\bm{\kappa},\bm{\tau},\bm{\beta})= \sum_{i=1}^{n} [ \, \delta_i \, \log \, h(y_i;\bm{x_i},\bm{\kappa},\bm{\tau},\bm{\beta}) + \log S(y_i;\bm{x_i},\bm{\kappa},\bm{\tau},\bm{\beta})],
\label{loglik}
\end{equation}
where $y_i$ are the time to event observations, $\delta_i$ is the censoring and $\bm{x_i}$ is the vector with the covariate information for the $i$-th patient.  The triples $ \{(y_1, \bm{x_1}, \delta_1),(y_2, \bm{x_2},\delta_2), ...,(y_n, \bm{x_n}, \delta_n) \}$ denote the random sample $ ( Y_i, \bm{X_i}, \Delta_i )$ of the $i$-th individual. Of course, the piecewise exponential model arises as special case when all $\kappa$'s are equal to 1. 

PWMs have been used at the past. There are some applications with known change-points \citep[see for example,][]{terawaki2006development}.
 A Bayesian approach to estimate the change-points was proposed in \cite{casellas2007bayesian}. In \cite{qian2014multiple} a likelihood-ratio test (LRT) statistic was proposed that can be used to test whether there is an abrupt change at an unknown point in the covariate coefficient vector in the Weibull hazard function. Similar theoretical results on the LRT are provided in \cite{li2014likelihood}.
For the PEM one can also see the work in \cite{dupuy2009detecting}. Hitherto, the works mainly attempt to join time intervals with common hazard. In the present paper we attempt to estimate the change-points together with the parameters of the model.

\subsection{Computational Details}
Maximizing the log-likelihood can be demanding due to the change-points since the log-likelihood surface can have jumps. To avoid such problems we propose Simulated Annealing as
the most appropriate derivative free method to optimize the
log-likelihood function $\ell(\bm{\theta})$, estimate the unknown
transition times and compare the efficiency of the treatments, for
both models. It is a generic probabilistic heuristic approach to
global optimization problems, investigated by
\citet{kirkpatrick1983optimization}. It locates a good
approximation to the global optimum in a large search space with
reasonable probability. This means that this optimization
algorithm allows moving to worse values of the objective function,
utilizing a probability, which decreases exponentially as the time
passes and as more iterations are performed. Accepting worse
solutions is a fundamental property of metaheuristics because it
allows a more extensive search for the optimal solution and
improves the obstacles of Local search algorithms.

The SA Algorithm \ref{alg:SAalgorithm} includes a parameter which is called \textit{temperature}. The SA is based on the Metropolis-Hastings algorithm and simulates the cooling process by gradually lowering the temperature of the system until it converges to a steady, frozen state \cite{chibante2010simulated}.
\begin{algorithm}[H]
\caption{Simulated Annealing algorithm \label{alg:SAalgorithm}}
\begin{algorithmic}[1]
\Require{ \begin{itemize} \item[] \item Select initial points for
the parameters; $\bm{\theta_0=(\kappa_0, \tau_0,
\beta_0)}$ as a random solution. \item Select initial and final
temperatures, let $T_0=500$ and $T_{\infty} > 0$. The temperature is the
Local Variable which controls the probability of downward steps.
\item Select the function $T_{r+1}=\frac{T_{r}}{1+0.01T_{r}}$ in which the
temperature decreases, $T_r$ denotes the temperature at iteration $r$. 
\end{itemize}
}
\\ 
At the $r$-th iteration, given the current value $\bm{ \theta}_r$, propose new points $\bm{\theta^*=(\kappa^*, \tau^*, \beta^*)}$ by sampling from a $N(\bm{ \theta}_r, \Sigma)$ where $\Sigma$ is diagonal with the same variance $\sigma^2$. 
\\
\vspace*{0.15cm} If  $\ell(\bm{\theta}^*) \ge \ell(\bm{\theta}_r)$, then
set $\bm{\theta}_{r+1}=\bm{\theta}^*$ and continue.
    Otherwise,
    \begin{itemize}
     \item Sample $u \sim U(0,1)$  \space \space 
     \item If $u < \exp \left( \frac{\ell(\bm{\theta}^*) - \ell(\bm{\theta}_r)}{T_r} \right)$ then $\bm{\theta}_{r+1}:=\bm{\theta}^*$ otherwise $\bm{\theta}_{r+1}:=\bm{\theta}_r$
\end{itemize}
Note that $\ell(\cdot)$ is given in \eqref{loglik}.
\\
Reduce the temperature $T$ according to the selected function
(cooling rate).
\\
Stop when you cannot find any better solution after a large number
of iterations.
\end{algorithmic}
\end{algorithm}

Some important details are:

\begin{itemize}
 \item Simulated annealing injects just the right amount of randomness into things to avoid the trapping attraction of local optimum early in the process without getting off course late in the game, when a solution is nearby. This makes it pretty good at tracking down a decent answer, no matter its starting point.
 \item For better optimization, when initializing the temperature variable we should select a temperature that will initially allow for practically any move against the current solution. This gives the algorithm the ability to better explore the entire search space before cooling and settling in a more focused region.
 \item The algorithm converges asymptotically to globally optimum solutions after a series of iterations and it does not allow any move against the optimum.
 \item  The initial temperature must be large enough to make the uphill and downhill transition probabilities about the same, which results in avoiding local optimums. On the other hand, the initial temperature must be such that not moving off the global optimum. Consequently, the definition of some control parameters (initial temperature, cooling rate, etc.) constitutes the main disadvantage, that is usual in local search algorithms, because it is subjective and must be based on empirical evidence \cite{kirkpatrick1983optimization}.
\end{itemize}

Standard errors of the estimated parameters can be obtained with standard bootstrap approach.

\section{Application} 
\subsection{About the Data} 
The data derived from a randomized clinical trial to investigate
the therapeutic effects of Acyclovir and Valaciclovir used also in
\cite{beutner1995valaciclovir}  and \cite{desmond2002clinical}. Overall, 1141 herpes zoster patients, aged 49 years or older, were randomized to orally
receive three different dosing schemes of treatment (see Table \ref{tab:demogr}). 
The three arms refered as 
ACV-7days (Acyclovir, 800mg, 5 per day for 7 days),
VACV-7days (Valaciclovir, 1000mg, 3 per day for 7 days) and 
VACV-14days (Valaciclovir, 1000mg, 3 per day for 14 days). 
The primary
clinical outcome is the  time to
complete cessation of ZAP (in days). Other demographic
characteristics, such as age and sex, are also included in the
dataset and can be seen in Table \ref{tab:demogr}. Three
observations were excluded since they reported no pain. All
statistical hypothesis tests have been conducted in the typical level of
significance 5\% and the analyses were carried out with the R 4.0.5  software package.
%and the computer programs are publicly accessible %in\url{https://github.com/Dimitraele/Piecewise-Survival-Models-HZ-virus}..

Table \ref{tab:demogr} provides some descriptive statistics about the data. %According to Table \ref{tab:demogr}, 
%the mean age of patients is approximately 68 %years and the mean time until complete pain %cessation is 60.67 days. Moreover, females and %white race are the categories which comprise %the majority of patients with percentages %56.79\% for females and 94.65\% for the white %race, while the most of them ($>90\%$) have %declared pain at baseline. 
%In Table \ref{tab:trt}, the percentages of %patients administered the three antiviral %schemes for the resolution of %zoster-associated pain are kept in balance. 
The herpes zoster data consist of 779 (68.45\%) events of completed pain cessation and 359 (31.55\%) censored observations.
%, as shown in Table \ref{tab:trtcens}. 
The majority of study population claimed pain duration before the 75th day of observation.
%(Figure \ref{fig:histZAP}). 
%It must be mentioned that the methodology of %\citet{desmond2002clinical} differs from the %present one, hence the existence of slight %discrepancies is absolutely reasonable. % --> %HERE OR IN THE DISCUSSION SECTION?

% in total ? too much?
% should we say about the enrollment of 49years old? like desmond?

\begin{table}[ht]
\footnotesize
\centering
\renewcommand{\arraystretch}{1.3}% Spread rows out...
\begin{tabular}{lclclclclclc}
\Xhline{1.2pt}
Characteristic & ACV, 7-day & \: \:  VACV, 7-day & \: \:  VACV, 14-day &  \: \: \space  Total \\
\space & n \space \space \space \; (\%)    & \; \; \;  n  \space \space \: \;(\%) & n  \space \space \space \; (\%) & \; \space N \space \space \space \; (\%) \\
\hline 
\hline
$n$ & 376 \: (32.95) &  384 \: 33.66 & 381 \: (33.39) & 1141\\
Events & 248 \: (31.84) & 263 \: (33.76) & 268 \: (34.40) &  779 \\
\hline 
\textbf{Mean (Range)}  \\
% \: \: ZTIMETO{\small *}  &  \: 68.17 \: (0, 172)   &  \;  56.26 \: (0, 172)  &  \; 57.73 \: %(0, 174) &  60.67 \: (0, 174) 
%\\
 \: \: Age, yrs   & \: 68.14 \: (50, 88) & \;  68.41 \: (49, 89)  & \;  67.78 \: (50, 89) &  68.11 \: (49, 89) \\
\hline
\textbf{Gender}  \\
 \: \: Female  & 232 \: (61.70) & \; \; 229 \: (59.64) & \: 187 \: (49.08) & \; 648 \: (56.79)\\
 \: \: Male    & 144 \: (38.30) & \; \; 155 \: (40.36) & \: 194 \: (50.92) & \; 493 \: (43.21)\\
\hline
\textbf{Race}  \\
\: \: White  & 354 \: (94.14) & \; \; 362 \: (94.27) & \: 364 \: (95.54) & \; 1080 \: (94.65)\\
\: \: Black  &  12 \: (3.20)  & \; \; \: 13  \: (3.39) & \: 10 \: (2.62) & \; \; \: 35 \: (3.07)\\
\: \: Other  & 10 \: (2.66) & \; \; \; \: 9 \: (2.34) & \: \: 7 \: (1.84) & \; \; \: 26 \: (2.28)\\
\hline
\textbf{Pain at baseline}  \\
\: \: Yes  & 338 \: (90.37) & \; \; 339 \: (88.74) & \: 348 \: (91.58) & \; 1025 \: (90.23)\\
\: \: No  & 36 \: (9.63) & \; \; \; 43 \: (11.26) & \: 32 \: (8.42) & \;\;\; 111 \: (9.77)\\
\: \: missing values  & 2 \; \; \; \; \;   & \; \; \; \; 2  &  1 \; \; \; \; & \;\;\; \; \: 5 \\
\Xhline{1.2pt}
\end{tabular} 
\\[0.15 em]
\vspace*{-0.15cm}
\begin{tablenotes}
\small
\item * Time in days to complete cessation of ZAP; 3 missing values \\
\end{tablenotes}
\caption{{\footnotesize Number of patients per arm and their demographic characteristics. Three observations were finally excluded from the analysis (one from each arm).}}
\label{tab:demogr}
\end{table}

\begin{center}
\begin{figure}[ht]
\includegraphics[scale=0.70]{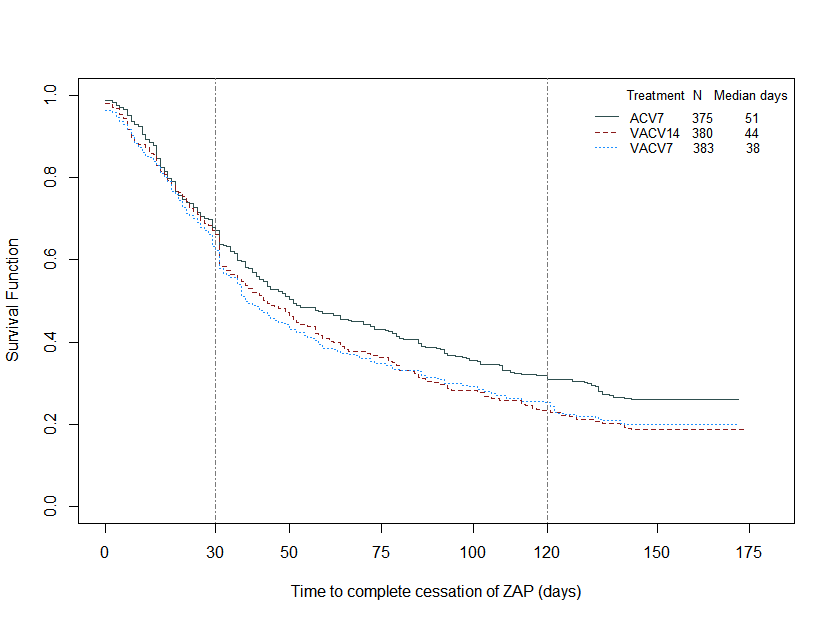}
\caption{\footnotesize {
Kaplan-Meier survival estimates for time to cessation of ZAP in patients with herpes zoster. }}
\label{fig:KM estimates}
\end{figure}
\end{center}

In Figure \ref{fig:KM estimates}, the patients administered Valaciclovir for 7 and 14 days seem to have similar behavior in their survival probabilities, which are considerably close to each other. This observation is further supported by the log-rank tests in Table \ref{tab:gehan-wilc}, which shows that K-M curves among the ACV, VACV7 and VACV14 treatment groups are statistically equivalent (\textit{p}-value= 0.08), while the therapies with ACV and merged VACV have significantly different K-M survival curves (\textit{p}-value= 0.03). This means that the two dosing schemes of VACV are not significantly different with respect to the resolution of herpes zoster-associated pain. Therefore, it is preferred to continue using the merged VACV treatment. Also, note that there is a sudden drop of survival probabilities close to the day 30, which gives the impression of a possible change in survival. 

Applying the Cox proportional hazard (PH) models in both cases for \textit{i.} ACV vs VACV7 vs VACV14 and \textit{ii.} ACV vs VACV result that Valaciclovir for 7 and 14 days significantly accelerated the ZAP resolution (\textit{p}-value=  0.0277 and \textit{p}-value=  0.037) compared to ACV, respectively. Specifically, complete cessation of pain occurs 20.7\% faster with Valaciclovir than Acyclovir \textit{p}-value=  0.0147), as shown in Table \ref{tab:HRstata}. 
%\textbf{According to the Wald tests, the PH assumption holds if there is no %significant time-varying covariate effect (\textit{p}-value=  0.971  and %\textit{p}-value= 0.634) for both cases.}  

\subsection{Piecewise Weibull Model}

The objective in this section is to explore the benefits of a parametric analysis of the ZAP data by using a piecewise Weibull distribution to model the times to complete cessation of ZAP. Two of these benefits are, firstly, to develop a data-driven approach to estimate the time-points where the hazard function changes, and, secondly,  to evaluate the hazard rates via an exhaustive local search algorithm. Finally, bootstrap approach was employed in order to obtain standard errors and 95\% confidence intervals of the estimates. Both approaches of known $(\tau_1=30, \tau_2=120)$ and unknown change-points are examined. 

\subsubsection{Known time change-points}

The log-likelihood for the PWM is used to test treatment effects within each phase of ZAP under the consideration of known transition times $(\tau_1=30, \tau_2=120)$ \cite{dworkin1994proposed} and estimated shape parameter $\bm{\hat{\kappa}}=(0.90,0.89,0.93)$. We compare the effectiveness of Acyclovir at 800mg five times a day for 7 days and Valaciclovir at 1000mg three times a day for 7 and 14 days in a clinical trial. We set $X_i=0$ if the $i$-th patient was treated with Acyclovir, and $X_i=1$ if Valaciclovir is administered to the $i$-th patient. The hazard rate of the \textit{i}-th patient in the \textit{j}-th phase is given by

\begin{equation}
 h_{ij}= \kappa_j \; e^{-(\beta{_j}^{ACV}+ \, \beta_{j}^{VACV} X_i)} \; t_i^{\kappa_j-1}, \;\;\;\;  i=1,\dots, n, \;\;\;\;\;  j=1,2,3 \, .
 \label{h2}
\end{equation}
In PWM case, the interest lies on the instantaneous hazard rate, which is computed individually for each patient due to its dependence on the observed time. The hazard rates for each treatment group and each phase have also been calculated at three indicative time points, $t$, which are tabulated in Table \ref{tab:pwmknown}. The hazard rates of pain resolution over time for Acyclovir and Valaciclovir, respectively, have a downward trend of: (i) 0.0136 to 0.0114 and 0.0153 to 0.0128 in acute phase; (ii) 0.0099 to 0.0087 and 0.0134 to 0.0117 in subacute phase; (iii) 0.00463 to 0.00452 and 0.00546 to 0.00534 in chronic phase. There is an obvious downward trend in hazard rates over time for both treatments which tends to be eliminated by crossing to the next stages of the disease. Moreover, the decline of hazard rates from subacute to chronic pain (or phase of postherpetic neuralgia) is larger than the corresponding one from acute to subacute. Both treatment arms are again statistically significant in patients who came up against the middle stage of ZAP, (\textit{subacute:} $\mathit{p_{ACV}}<$ 0.0001; $\mathit{p_{VACV}}=$ 0.0087), while it is observed that patients who received Acyclovir have slightly smaller hazard rate. Testing the hypothesis $H_0 : \beta^{VACV}=0$ for each phase indicates statistically significant difference among the treatments for the subacute phase (see Table \ref{tab:pwmknown}). The subacute phase of pain lasts longer until ZAP resolution as shown in Figure \ref{fig:hazpwmknown}. Also note that the estimated shape parameters for the different Weibulls at each phase are  $\bm{\hat{\kappa}}=(0.90,0.89,0.93)$ and they are smaller than 1 as implied by an exponential survival model (constant hazards). A LRT statistic testing the piecewise exponential model against the piecewise Weibull (using the same known time points for change) has a value 14.62 with 3 degrees of freedom and a \textit{p}-value equal to 0.0022 justifying the choice of the piecewise Weibull model. 

\vspace{0.25cm}

\begin{table}[ht]
\centering
\footnotesize
 \renewcommand{\arraystretch}{1.3}% Spread rows out...
  \begin{tabular}{c c c c c c c c}  
\Xhline{1.2pt}
Phase & & &  Treatment & & & \\ \cline{2-7}
  & & Acyclovir & & &  Valaciclovir &\\
 t (days) & Estimate   & \textit{p}-value{\small{*}} & Hazard rate  & Estimate  & \textit{p}-value{\small{**}} & Hazard rate & Diff{\small{***}}\\
 & (s.e.)  & & estimate & (s.e.) & & estimate & {\tiny VACV-ACV}\\
\hline
Acute  & 4.030  & $<0.0001$ &  & -0.115  & $0.3144$ &  & \\
 &(0.0945) &  &  &  (0.1144) &  & & \\
 5  &   &  & 0.0136 &   &  & 0.0153 & \\
 14 &   &  & 0.0012 &   &  & 0.0014 & 0.0002 \\
 29 &   &  & 0.0114 &   &  & 0.0128 & \\
Subacute   & 4.098  & $<0.0001$ &  & -0.292  & $0.0087$ &  &\\
 &(0.0921) &  &  &  (0.111) &  &  &\\
 35 &   &  & 0.0099 &   &  & 0.0134  & \\
 58 &   &  & 0.0095 &   &  & 0.0127  & 0.0032 \\
119 &   &  & 0.0087 &   &  & 0.0117  & \\
Chronic   & 4.965  & $<0.0001$ &  & -0.165  & $0.578$ &  &\\
 &(0.2357) &  &  &  (0.2981) &  & &\\
 125 &  &  & 0.00463  &   &  & 0.00546  & \\
 158 &  &  & 0.00455  &   &  & 0.00537  & 0.0008\\
 174 &  &  & 0.00452  &   &  & 0.00534  & \\
\hline 
 & Loglikelihood: & -4233.693  &  & AIC: & 8479.386 &  &\\
\Xhline{1.2pt}
\end{tabular}
\\[0.15 em]
\vspace*{-0.15cm}
\begin{tablenotes}
\small
\item *  $H_0 : \beta^{ACV}=0$  \\
\item ** $H_0 : \beta^{VACV}=0$ \\
\item *** Diff = Hazard Rate$_{VACV}$ - Hazard Rate$_{ACV}$; The differences correspond to the mean time of each phase.
\end{tablenotes}
\caption{{\footnotesize ML estimates of the parameters of PWM and treatment comparison, assuming change-points on the 30th and 120th day of observation and the estimated shape parameters $\bm{\hat{\kappa}}=(0.90,0.89,0.93)$. Hazard rates are also calculated based on the hazard formula \eqref{h2} across the phases.}}
\label{tab:pwmknown}
\end{table}

\begin{figure}[ht]
\centering
\renewcommand{\arraystretch}{1.3}% Spread rows out...
\begin{subfigure}{.5\textwidth}
\raisebox{-0.9\height}{\includegraphics[scale=0.47]{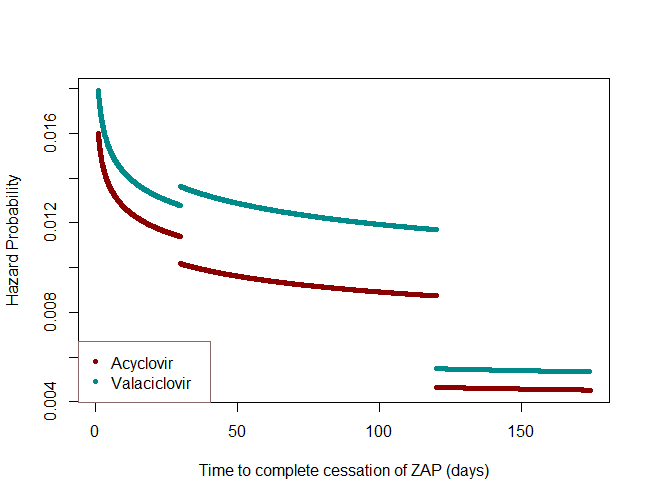}}
\caption{ }
\end{subfigure}%
\begin{subfigure}{.5\textwidth}
\raisebox{-0.9\height}{\includegraphics[scale=0.47]{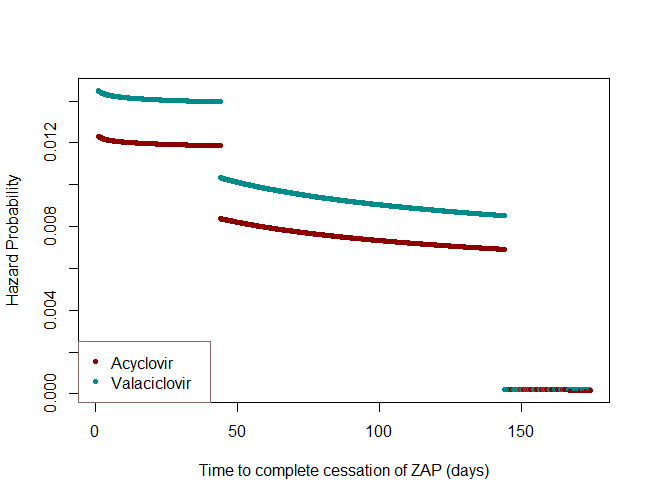}  %hazpwmunknown_new.png
} 
\caption{ }
\end{subfigure}%
\caption{{\footnotesize Hazard rate estimates based on PWM with: a. known change-points $\boldsymbol{\tau}=(30,120)$ and  $\bm{\hat{\kappa}}=(0.90,0.89,0.93)$; b. the SA average estimated change-points $\boldsymbol{\hat{\tau}}=(44,144)$ and $\bm{\hat{\kappa}}=(0.99,0.84,0.36)$.}}
\label{fig:hazpwmknown}
\end{figure}

\subsubsection{Unknown time change-points}

Now the change-points, i.e. the times the hazard rates change need to be estimated. 
The SA algorithm was applied to the herpes zoster data. Here we assumed that the two change-points were unknown and we left the SA to estimate them as two extra parameters of the model. Ten independent chains of the exhaustive local search algorithm (SA) of 6000 iterations were executed, setting the initial values ($\boldsymbol{\kappa_0}, \boldsymbol{\tau_0}, \boldsymbol{\beta_0})$ on the parameters of interest. In addition, we set $T_0 = 500 $ as an initial temperature, and the decreasing function was $T_{new}=\frac{T_{old}}{1+0.01T_{old}}$ \cite{lundy1986convergence}.
Some details about the implementation of the SA algorithm follow.
We used the 30th and 120th days as initial values for change-points, based on
classification in  \citet{dworkin1994proposed}. Nevertheless, these values were randomly varied with common variance $\sigma^2$ during the estimation procedure. Similarly, the initial values of the regression coefficients $\boldsymbol{\beta}$ (Equation \eqref{h2}) were estimated through a simple log-likelihood optimization process given the change-points $\boldsymbol{\tau}=(30,120)$. Consequently, the selection of initial values plays substantial role in the estimation with SA algorithm.

Table \ref{tab:pwmunknown} summarizes the results of 100 bootstrap replications of the SA algorithm including the parameter estimates (i.e. transition times, hazard rates and shape parameters) across the phases. The first change in ZAP hazard function is estimated at $ \hat{\tau_1}=44$ days on average (SE=9.48), while the second appears on average at $\hat{\tau_2}=144$ days (SE=3.06). There is a downward trend in hazard rate across the phases for both treatment arms. Moreover, there is still considerable difference in hazard rates among Acyclovir and Valaciclovir, but only over the first two phases (acute and subacute), while in the chronic phase there is no distinction between the two treatments (see Figure \ref{fig:hazpwmknown}). In agreement with the literature, this study of zoster-associated pain shows that Valaciclovir is more effective than Acyclovir in healing acute lesions. Recall that the commonly used values for the change-points were 30 and 120, respectively. The value of 30 lies inside the 95\% confidence interval for the first point (interval lies from (24,58)), while for the second one 120 is perhaps too early (interval lies from (137,149)). Also note that for the shape parameters the value 1 is included in the confidence interval corresponding to the first phase, marginally not inside the confidence interval of the second phase, while for the third phase the interval boundaries are far from 1 implying that the constant hazards assumption is not valid overall.
A final point relates to the second change-point which appears too late for the data. After this time point we have not observed any other event.

It is also interesting to notice that the standard error of the first change-point $(\widehat{\tau}_1)$ is larger than the standard error of the second $(\widehat{\tau}_2)$, as given in Table \ref{tab:pwmunknown}. This might happen because most patients with herpes zoster declared pain duration before the second phase, as shown in Figure \ref{fig:KM estimates}, and less events observed during the second and the third phases.

\comm{
\begin{figure}[htbp]
\centering
\renewcommand{\arraystretch}{1.3}% Spread rows out...
\caption{SA Hazard rate estimates based on the piecewise Weibull model using the estimated change-point vector and the shape parameters ; \;\;\;\;\;\;\;\;\;\;\;\;\;\;\;\;\;\;\;\;\;\;\;\;\;\;\;\;\;\;\;\;\;\;\;\;\;\;\;\;\;\;\;\;\;\;\;\; \textit{(i)} $\boldsymbol{\tau}=(33,118)$ , \textit{(ii)} $\boldsymbol{\kappa'}=(0.851,0.871,0.929)$.}
\begin{tabular}{clclcl}
  \raisebox{-0.9\height}{\includegraphics[height=9.7cm]{../figures/hazz.png}} &
\end{tabular}
\end{figure}
}

Figure \ref{fig:cumhaz_pwm} illustrates the parametric piecewise Weibull cumulative hazard estimates for both treatments. There is considerable difference when comparing the fitted cumulative curve slopes from PWM between the proposed change-points $\boldsymbol{\widehat{\tau}_{p}}=(30,120)$ \cite{dworkin1994proposed} and the expected estimated change-points $(44,144)$ as well as the existence of the change-points is more evident in the right figure. Akaike's criterion (AIC) was implemented to compare the candidate models regarding the analysis of ZAP data. According to the Table \ref{tab:AICs}, there is significant fit difference among the piecewise exponential and piecewise Weibull modelling. In fact, the 4th model (that is, PWM with change-points at 44 and 144 days) seems to be the selected one. Also note that from Figure \ref{fig:hazpwmknown} one can see that when estimating the change-points later than the 30 days, this has a large effect on the changes in the hazard rates.

Interest lies on the graphical presentation of the log-likelihood given each candidate pair of change-points and its behavior with respect to $\tau_1$ and $\tau_2$, as shown in Figure \ref{fig:Loglike2}. One can see that the log-likelihood is relatively flat around the maximum value indicating the increased uncertainty around the change-points as already noted. 

\begin{table}[H]
\centering
\footnotesize 
\renewcommand{\arraystretch}{1.3}% Spread rows out...
\begin{center}
\begin{tabular} {c c c c c} 
\Xhline{1.2pt}
  & &  \textbf{Change-points} & \\
 \hline
 SA Estimate & Mean  &  Standard error  &  $95\% \: CI^*_{ \widehat{\tau_1}}$ &  $95\% \: CI^*_{ \widehat{\tau_2}}$\\
 $(\widehat{\tau_1},\widehat{\tau_2})$ & $E(\widehat{\boldsymbol{\tau}})$ &$(SE_{\widehat{\tau_1}},SE_{\widehat{\tau_2}})$  & & \\
 \hline
 \hline
(46,143) & (44,144) & (9.48,3.06)  & (24,58) & (137,149)  \\
%(43,130) & (44,137) & (9.26,10.77) & (1,7) & (32,65) & (122,157)  \\
\Xhline{1.2pt}
 & & & \\
\Xhline{1.2pt}
  & &  \textbf{Shape parameters} & \\
 \hline
Parameter & SA Estimate & Mean  &  Standard error  &  $95\% \: CI^*$  \\
 \hline
 \hline
$\kappa_1$     & 1.05 & 0.99  & 0.058   & (0.907,1.139)  \\
$\kappa_2$     & 0.90 & 0.84  & 0.096   & (0.70,0.99)  \\
$\kappa_3$     & 0.37 & 0.36  & 0.172   & (0.112,0.676)  \\
\Xhline{1.2pt}
 & & & \\
\Xhline{1.2pt}
& &  \textbf{Hazard rates} & \\
& &  \textbf{for ACV} & \\
\hline
Parameter \; ($t^{**}$) & SA Estimate & Mean & Standard error  & $95\% CI^*$ \\
\hline 
\hline
$h_{1}^{(0)} \; (14)$   & 0.0130 & 0.0140 & 0.0029  & (0.009,0.020)  \\
$h_{2}^{(0)} \; (58)$   & 0.0069 & 0.0092 & 0.0049  & (0.003,0.019)  \\
$h_{3}^{(0)} \; (158)$  & 0.0002 & 0.0002 & 0.0001 &  (0.0001,0.0006)  \\
\Xhline{1.2pt}
 & & & \\
\Xhline{1.2pt}
& &  \textbf{Hazard rates} & \\
& &  \textbf{for VACV} & \\
\hline
Parameter \; ($t^{**}$) & SA Estimate & Mean & Standard error & $95\% CI^*$ \\
\hline 
\hline
$h_{1}^{(1)} \; (14)$   & 0.0159 & 0.0161 & 0.0028  & (0.011,0.021)  \\
$h_{2}^{(1)}\; (58)$    & 0.0110 & 0.0115 & 0.0063  & (0.004,0.024)  \\
$h_{3}^{(1)}\; (158)$   & 0.0003 & 0.0002 & 0.0002  & (0.00007,0.0006)  \\
\hline
Loglikelihood:& -4189.30 & & &  \\
\Xhline{1.2pt}
\end{tabular}
\end{center}
\vspace*{-0.3cm}
\begin{tablenotes}
\small
\item * Percentile confidence intervals were computed.
\item ** Hazard rate $h_j^{(k)}$ corresponds to the $k$-th treatment at phase $j$ evaluated at the specific time $t$. 
\end{tablenotes}
\caption{{\footnotesize Simulated Annealing parameter estimates, standard errors and 95\% CI for both treatments Acyclovir (ACV) and Valaciclovir (VACV) derived by 100 nonparametric Bootstrap replications.}}
\label{tab:pwmunknown}
\end{table}

\begin{center}
\begin{figure}[H]
\centering
\renewcommand{\arraystretch}{1.3}% Spread rows out...
%\vspace*{0.2cm}
\begin{subfigure}{.5\textwidth}
\raisebox{-.7\height}{\includegraphics[height=6cm]{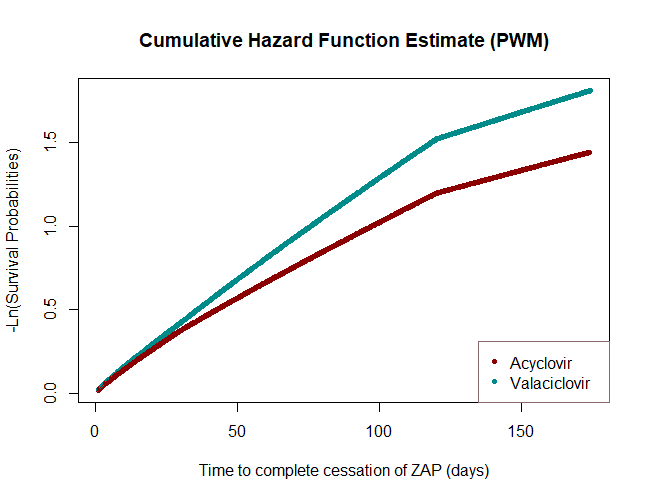}}   
\caption{ }
\end{subfigure}%
\begin{subfigure}{.5\textwidth}
\raisebox{-0.7\height}{\includegraphics[height=6cm]{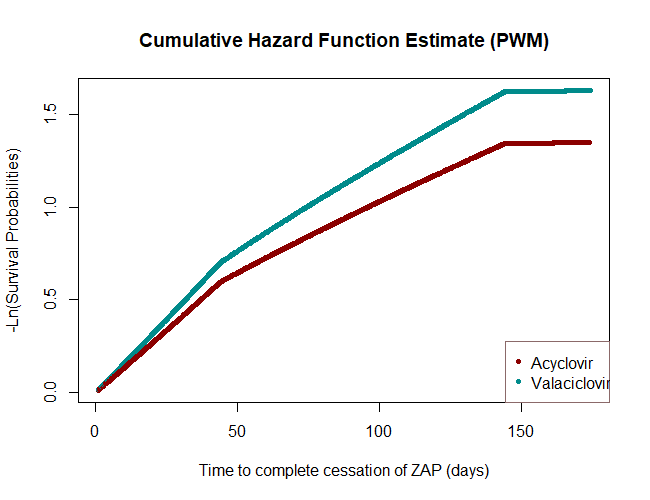}}   
\caption{ }
\end{subfigure}%
\caption{Fitted Cumulative Hazard Estimate of PWM based on a. the proposed change-points (30,120), and b. the average estimated change-points (44,144). }
\label{fig:cumhaz_pwm}
\end{figure}
\end{center}

\comm{
\begin{center}
\begin{figure}[htbp]
\centering
\renewcommand{\arraystretch}{1.3}% Spread rows out...
\caption{Predicted survival of PEM and PWM VS Kaplan Meier survival estimates based on the expected estimated change-points a. at (46,142), and b. at (44,144). } 
\begin{subfigure}{.5\textwidth}
\raisebox{-.7\height}{\includegraphics[height=6.5cm]{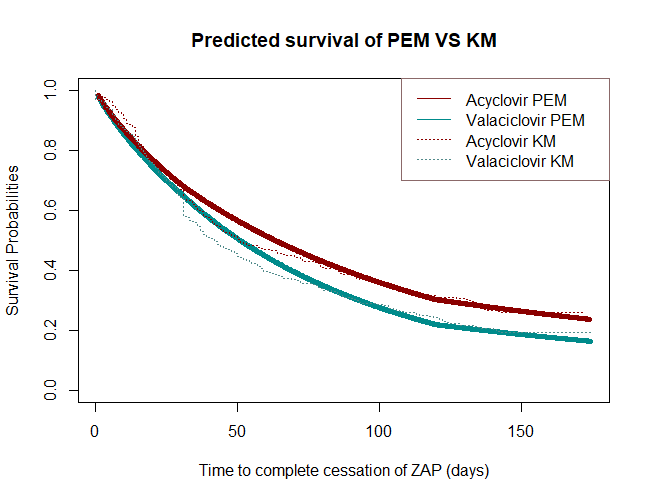}}  
\caption{ }
\end{subfigure}%
\begin{subfigure}{.5\textwidth}
\raisebox{-0.7\height}{\includegraphics[height=6.5cm]{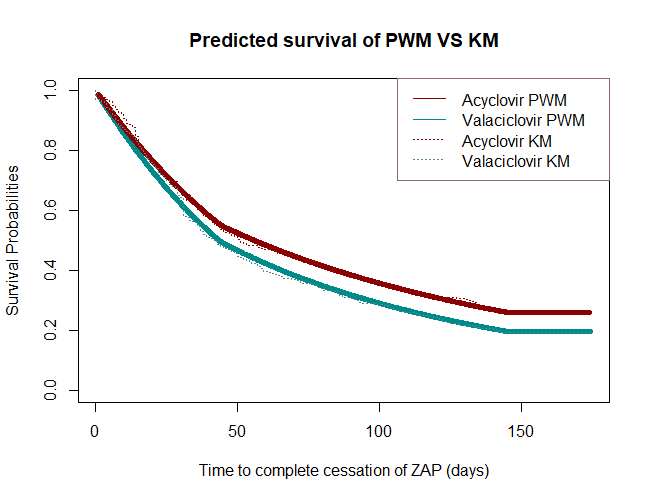}}     
\caption{ }
\end{subfigure}%
\end{figure}
\end{center}
}

\begin{table}[htbp] %[htp]
\centering
\small
 \renewcommand{\arraystretch}{1.3}% Spread rows out...
\begin{center}
\vspace*{-0.35cm}
  \begin{tabular}{lclclclclc}   
\Xhline{1.2pt}
 Model &  change-point pair & $p^{*}$ & Loglikelihood & $AIC^{**}$ \\
\hline 
\hline
PEM   & (30,120) & 6 & -4241 &     8494\\
      & (46,142) & 8 & -4206.22 &  8428.44  \\ 
\hline
PWM   & (30,120) & 9 & -4233.693 &  8485.39 \\  
      & (44,144) & 11 & -4189.30  & 8400.6\\  
\Xhline{1.2pt}
\end{tabular}
\end{center}
\vspace*{-0.60cm}
\begin{tablenotes}
\centering
\small
\item * $p$ denotes the number of model parameters
\item ** $\text{AIC} = -2\log( \text{likelihood}) + 2p $
\end{tablenotes}
\caption{SA goodness of fit statistics of 4 different piecewise regression models, assuming 2 transition times in hazard function.}
\label{tab:AICs}
\end{table}

\begin{figure}[htbp] %[htbp]
\centering
\renewcommand{\arraystretch}{1.3}% Spread rows out...
\begin{tabular}{clclcl}
  \raisebox{-0.9\height}{\includegraphics[scale=0.8]{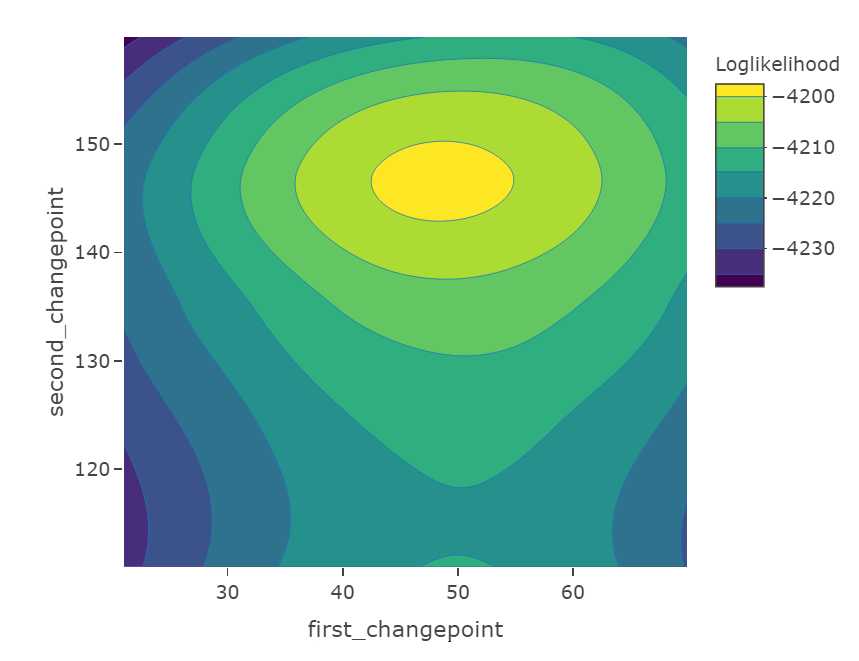}} &  
\end{tabular}
\caption{{\footnotesize Visualization of the log-likelihood behavior given the candidate change-point pairs.}}
\label{fig:Loglike2}
\end{figure}

\justify
\subsection{Equality of hazards across the phases} 
\justify
%% question 2 in Arani? pg 2431
%% pg 2432 arani see paragraph +++
The equality of hazard rates across the distinct phases constitutes an important part of the change-point analysis because it verifies the number of phases. Thus, modelling can be improved concerning the goodness of fit and complexity of the model. Based on the hypothesis testing of \citet{arani2001phase}, the comparison of hazards from the best fitted model; i.e. the PWM model with  change-points at (44,144) can be tested as
%$$H_0: \frac{\lambda_j}{\lambda_i}=1 ,\;\;\; \text{where} \;\;\; i<j \;\;\; \text{and} \;\;\; i,j=1,2,3.$$
$$H_0: \frac{h_j}{h_i}  =1 ,\;\;\; \text{where} \;\;\; i<j \;\;\; \text{and} \;\;\; i,j=1,2,3.$$
This means that under the null hypothesis the hazard rate of herpes zoster pain resolution in phase \textit{i} is equal to the corresponding one of phase \textit{j}. As shown in Table \ref{tab:comparison}, the hazard rates of pain cessation are significantly different between all phases for both treatment groups.

\begin{table}[htbp] %[htbp]
\centering
\small
\renewcommand{\arraystretch}{1.3}% Spread rows out...
\begin{center}
\begin{tabular} {c c c} 
\Xhline{1.2pt}
  &   \textbf{Acyclovir} & \\
 \hline
 Parameter & Ratio Estimate  &  $\text{\textit{p}-value}^*$ \\
 \hline
 \hline
$h_2^{(0)}/h_1^{(0)}$     & 0.598  & $<0.0001$ \\
$h_{3}^{(0)}/h_{1}^{(0)}$     & 0.029  & $<0.0001$ \\
$h_{3}^{(0)}/h_{2}^{(0)}$     & 0.048  & $<0.0001$  \\
\Xhline{1.2pt}
 & & \\
\Xhline{1.2pt}
&   \textbf{Valaciclovir} & \\
\hline
Parameter & Ratio Estimate  &  $\text{\textit{p}-value}^*$ \\
\hline 
\hline
$h_{2}^{(1)}/h_{1}^{(1)}$    & 0.627 & $<0.0001$   \\
$h_{3}^{(1)}/h_{1}^{(1)}$    & 0.025 & $<0.0001$   \\
$h_{3}^{(1)}/h_{2}^{(1)}$    & 0.040 & $<0.0001$   \\
\Xhline{1.2pt}
\end{tabular}
\end{center}
\vspace*{-0.2cm}
\begin{tablenotes}
\small
\item * \textit{p}-values correspond to $H_0: \frac{h_j}{h_i}=1 ,\;\;\; \text{where} \;\;\; i<j \;\;\;\text{and} \;\;\; i,j=1,2,3$.
\item ** Hazard rates correspond to the mean time of each phase.
\end{tablenotes}
\caption{Ratios of hazard estimates of $j=1,2,3$ phases for treatments Acyclovir ($k=0$) and Valaciclovir ($k=1$). Here  $h_j^{(k)}$ denotes the hazard rate for treatment $k$ at phase $j$. The testing procedure is the same as in \citet{arani2001phase} }
\label{tab:comparison}
\end{table}

%\subsection{Computational Remarks} 
% There is no doubt with regards to the validity of the Simulated Annealing %estimation method since the standard errors estimated by bootstrap techniques %are fairly small. Furthermore, the main reason of defining 30th and 120th %days as initial values for change-points came from the classification in  %\citet{dworkin1994proposed}. Nevertheless, these values were randomly varied %with common variance $\sigma^2$ during the estimation procedure. Similarly, %the initial values of the regression coefficients $\boldsymbol{\beta}$ %(Equation \eqref{h2}) were estimated through a simple log-likelihood %optimization process given the change-points $\boldsymbol{\tau}=(30,120)$. %Consequently, the selection of initial values plays substantial role in the %estimation with SA algorithm. 

%\textbf{Simulated Annealing in R.} In addition, the code of Simulated Annealing was %constructed for local search applications. The usefulness of the algorithm lies on the %fact that it provides flexibility regarding the detection of multiples change-points %detection. The SA estimation process is based on the convergence of the maximum %likelihood and its major advantage is that not only detects the change-points, but 5simultaneously produces the optimum solutions of other model parameters (i.e.  hazard %rates and shape parameters) in each piece of the likelihood function. 

\section{Conclusions}
%Herpes Zoster is an infection that results when varicella-zoster virus (VZV) reactivates from its latent state. Thus, Herpes Zoster can be associated with a variety of neurologic complications. Symptoms usually begin with pain and vesicular eruption along the affected skin. Pain resolution is characterized by three phases: \textit{acute} which is marked by dermatomal vesicles and pain, \textit{subacute} and \textit{chronic}. Posthereptic neuralgia (PHN) is usually included into subacute and chronic phases appearing 4 weeks after the onset of lesions regardless of when the lesions resolved. Prompt antiviral regiments for ZAP, such as Acyclovir and Valaciclovir, can accelerate rash healing, reduce pain duration and perhaps the incidence of postherpetic neuralgia % wood2002understanding, robert? [1,2].
%\\\\

In this paper we revisited a data set related to Herpes Zoster
pain using the piecewise Weibull model that offers certain
advantages and flexibility while keeping the idea of the three
phases of the disease pain. The model has also covariates and it
does not assume constant hazard for each stage and thus allows for
different behaviors.

 We have implemented an approach where the change-points of the piecewise model are considered as unknown parameters and
we need to estimate them. This can also be the basis to confirm
the theoretical part about the pain phases. To avoid numerical
problems due to the change-points we implemented a
 Simulated Annealing algorithm to estimate simultaneously the
 change-points and the parameters at each stage.
 %Simulation
 %evidence (not reported here) supports that the method works fine.

An interesting question related to how many change-points exist
is not treated here in detail. It seems that LRT approaches can be
useful for that, especially since the SA method allows for fitting
the model without problems. In our case, working with more than 2 changing points (3 phases alternatively) we did not find significant improvements in the log-likelihood.

The proposed piecewise Weibull model, accompanied by the Simulated Annealing estimation algorithm, presents an improved method to analyse ZAP data by incorporating in the modelling the covariate information and a variety of other model parameters in each phase. Based on the paper results, the piecewise Weibull model demonstrated its superiority over the commonly used piecewise exponential modelling.
  As far as the disease related findings, the hazard function is almost flat over the acute phase, one may think that applying a mixture of piecewise regression models such as a piecewise exponential model for the acute phase and and a piecewise Weibull model for the subacute and chronic phases could possibly lead to a better fit. Finally, given that Gnann and Whitley \citet{gnann2002clinical} have introduced a prodromal phase lasting the 1-5 days before the rash onset, the extension of the proposed model to accommodate the number of change-points as an additional model parameter will increase the flexibility of the proposed model and it might improve further its accuracy.

\section{Acknowledgements}
We are grateful to  GlaxoSmithKline Research \& Development Ltd for sharing their laboratory datasets obtained through www.ClinicalStudyDataRequest.com to carry out this research work. The data have been collected following all the appropriate ethical approval procedures. We would also like to thank the Laboratory of Bayesian and Computational Statistics of AUEB for giving access to its resources. This research received no grant from any funding agency or from commercial or not-for-profit sectors. The authors declare that there are no conflicts of interest. 

\medskip
{\small \bibliography{Bibliography} } 
% The references (bibliography) information are stored in the file named "Bibliography.bib"

\newpage
\begin{appendix}
\vspace*{-0.3cm}
\subsection*{Appendix - Tables}

\vspace*{-0.3cm}

\setcounter{table}{0}
\renewcommand{\thetable}{A.\arabic{table}}
\renewcommand{\theHtable}{A.\thetable}

\begin{table}[htbp]
\centering
\small
\renewcommand{\arraystretch}{1.3}% Spread rows out...
\begin{tabular} {c c c c c c } 
\Xhline{1.2pt}
  & N & Observed & Expected & (O-E)$^2$/ E & (O-E)$^2$/ V \\
\hline 
\hline
Acyclovir 7 days     & 375 & 153 & 172 & 2.17 & 4.847 \\
Valaciclovir 7 days  & 383 & 168 & 155 & 1.08 & 2.278 \\
Valaciclovir 14 days & 380 & 166 & 160 & 0.25 & 0.535 \\
\hline
\; X$^2$statistic = 5.1 & \;\;\; df = 2,  & \;\;  \text{\textit{p}}-value = 0.08 & & & \\
\hline 
\hline
Acyclovir      & 375 & 153 & 172 & 2.17 & 4.85 \\
Valaciclovir   & 763 & 335 & 315 & 1.18 & 4.85 \\
\hline
\; X$^2$statistic = 4.8 & \;\;\; df = 1,  & \;\;  \text{\textit{p}-value} = 0.03 & & & \\
\Xhline{1.2pt}
\end{tabular}
\caption{Gehan-Wilcoxon logrank test for comparing survival curves between treatment groups: \textit{(i)} Acyclovir for 7 days, Valaciclovir for 7 days, Valaciclovir for 14 days, and \textit{(ii)} Acyclovir for 7 days, Valaciclovir for 7 and 14 days.}
\label{tab:gehan-wilc}
\end{table}

\begin{table}[htbp]
\centering
\small

\comm{
\caption{Quantiles based on the Kaplan-Meier}
\renewcommand{\arraystretch}{1.3}% Spread rows out...
\begin{tabular} {c c c c} 
\Xhline{1.2pt}
 & $Q_{25}$ & $Q_{50} $ & $Q_{75}$ \\
\hline 
\hline
Acyclovir for 7 days      & 21 & 51 & NA  \\
Valaciclovir for 7 days   & 20 & 38 & 121 \\
Valaciclovir for 14 days  & 22 & 44 & 113 \\
\Xhline{1.2pt} 
\end{tabular}
}

%\bigskip
%\vspace*{1.7cm}

% Log likelihood 3 without tvc =   -4757.4387  
% Log likelihood 3 with tvc  =   -4750.8059 
\centering
\small
\renewcommand{\arraystretch}{1.3}% Spread rows out...
\begin{tabular} {ccccc} 
\Xhline{1.2pt} 
  No. of    &  &   &  & \\
 different doses & Variable  & HR & 95\% C.I. & \textit{p}-value\\
\hline 
\hline
 & Valaciclovir for 7 days vs Acyclovir   & 1.217        &(1.022, 1.450) & 0.0277*   \\ \vspace*{-0.5cm}
  & & & &  \\ 
 3 & Valaciclovir for 14 days  vs Acyclovir & 1.204 & (1.011, 1.433) & 0.0370*\\   \vspace*{-0.5cm}  
   & & & &  \\ 
 & Valaciclovir for 7 days  vs Valaciclovir for 14 days & 1.011 & (0.852, 1.201) & 0.898  \\
\hline
  2 & Valaciclovir vs Acyclovir   & 1.207        &(1.038, 1.403) & 0.0147* \\   
  \vspace*{-0.5cm}
    & & & &    \\     
\Xhline{1.2pt}
\end{tabular}
\begin{tablenotes}
\small
\item * $\text{\textit{p}-value} < 0.05$.
\end{tablenotes}
\caption{Hazard Ratios for ZAP cessation of Cox models considering three or two different dosing schemes for herpes zoster disease. }
\label{tab:HRstata}
\end{table}

\end{appendix}

\end{document}